\documentclass[runningheads]{llncs}
\usepackage[T1]{fontenc}
\usepackage{graphicx}
\usepackage{amsmath}
\usepackage{booktabs,collcell,eqparbox,makecell}
\usepackage{subcaption}
\usepackage{multirow}
\usepackage{algorithmic}

\begin{document}
\title{Analysis of Smooth Pursuit Assessment in Virtual Reality and Concussion Detection using BiLSTM}
\titlerunning{Analysis of Smooth Pursuit Assessment}

\author{Prithul Sarker\inst{1} \and
Khondker Fariha Hossain\inst{1} \and
Isayas Berhe Adhanom\inst{1} \and \\
Philip K Pavilionis\inst{2} \and
Nicholas G. Murray\inst{2} \and
Alireza Tavakkoli\inst{1}
} 

\authorrunning{Sarker et al.}

\institute{Department of Computer Science and Engineering, University of Nevada, Reno, United States \and 
School of Public Health, University of Nevada, Reno, United States
\email{prithulsarker@nevada.unr.edu}}


\maketitle              
\begin{abstract}
The sport-related concussion (SRC) battery relies heavily upon subjective symptom reporting in order to determine the diagnosis of a concussion. Unfortunately, athletes with SRC may return-to-play (RTP) too soon if they are untruthful of their symptoms. It is critical to provide accurate assessments that can overcome underreporting to prevent further injury. To lower the risk of injury, a more robust and precise method for detecting concussion is needed to produce reliable and objective results. In this paper, we propose a novel approach to detect SRC using long short-term memory (LSTM) recurrent neural network (RNN) architectures from oculomotor data. In particular, we propose a new error metric that incorporates mean squared error in different proportions. The experimental results on the smooth pursuit test of the VR-VOMS dataset suggest that the proposed approach can predict concussion symptoms with higher accuracy compared to symptom provocation on the vestibular ocular motor screening (VOMS).

\end{abstract}
\section{Introduction}

Mild Traumatic Brain Injuries (mTBI), also known as concussions, remain an active major public health issue that affects all levels of participation in sport and recreational activities with an average of 1.6 to 3.8 million occurrences each year \cite{daneshvar2011epidemiology}. A higher risk of traumatic brain injury (TBI) is linked with many of these activities. In particular TBI affects an estimated 1.7 million people every year in the United States according to the estimation of the US Centers for Disease Control and Prevention (CDC) \cite{daneshvar2011epidemiology}. Sport-related concussions (SRCs) are a variety of injuries that produce transitory neurological impairment and usually go away within two to four weeks \cite{kontos2021discriminative}.

Assessment of concussion is challenging as it requires recognizing the post-injury symptoms rapidly. Some of the most validated and utilized sports concussion assessment tools are Sport Concussion Assessment Tool 5 (SCAT5) \cite{echemendia2017sport}, Standard Assessment of Concussion (SAC) \cite{mccrea2003acute}, King-Devick test (KDT) \cite{galetta2016king}, Balance Error Scoring System (BESS) \cite{bell2011systematic}, Vestibular/Ocular Motor Screening Tool (VOMS) \cite{mucha2014brief}, Blood-based biomarkers \cite{shahim2014blood} etc. The tests often rely on the athlete’s self-reported symptoms, and sometimes the symptoms go unrecognized for the clinician perspective, resulting in misleading outcomes. Additionally, the physicians' concussion assessment system is sometimes influenced by self-reported bias \cite{yorke2017validity}. Even the most extensively used tests to detect concussion such as SCAT5 are not always accurate \cite{mckeithan2019sport}.

An immediate and transient loss of consciousness at the instant of head trauma and returning to alertness within an hour or two is the main symptom of concussion. Two of the commonly reported trajectories after SRC are 1) vestibular with dizziness, nausea, and vertigo as common symptoms and 2) ocular with convergence insufficiency, blur, diplopia and/or headaches as impairments \cite{kontos2021discriminative}. About 30\% of concussed athletes have reported visual impairments during the first week of the injury \cite{mucha2014brief}. The detection of vestibular and visual impairments is the component of VOMS assessment to concussion evaluation and management. The VOMS study contains five concise evaluations: (1) smooth pursuit, (2) horizontal and vertical saccades, (3) convergence, (4) vestibular ocular reflex (VOR), and (5) visual motion sensitivity (VMS). The test provokes the systems responsible for controlling balance, vision, and movement, which help identify issues that are not found in other assessments \cite{mucha2014brief}. 


In our study,  we used virtual reality (VR) headset to develop different domains of the assessment to eliminate subjectiveness and to introduce objectiveness in the VOMS assessment. Smooth pursuit of the VOMS assessment is performed to examine if there is a central pathology that prevents the eyes from tracking moving targets. 

\section{Related Work}

Concussion diagnosis is primarily justified in order to reduce the risk of immediate or delayed harm following an external force head injury.
As a result, tests must be able to spot deviations from normal performance, behavior, and symptoms as well as changes or risks.
The Standard Assessment of Concussion (SAC), developed in 1997, comprises measures 
Orientation, short-term memory, attention, and delayed recall are the four components of the Standard Assessment of Concussion (SAC), which assesses mental health~\cite{mccrea2003acute}.
Despite being a part of the first concussion evaluation, the SAC is insufficient on its own to conduct a complete concussion evaluation.
The SAC score cannot be used to determine the severity of a concussion or to support a player's return to play on its own~\cite{howell2015return}.

The Balance Error Scoring System (BESS) was developed by Guskiewicz et al. to assess an athlete's balance after a concussion~\cite{bell2011systematic}. Trials for the BESS test must be completed by the athlete.
A sequence of six balancing poses is used by BESS to test postural stability by subjectively counting mistakes.
The number of errors on each trial determines the BESS score~\cite{bell2011systematic}.
Although numerous studies now conclude that there are postural and motor signs and symptoms that do not go away after a week or two, the BESS still has a tendency to be most sensitive during the very acute period of recovery~\cite{baker2014visuomotor}\cite{howell2015return}.

The Sideline Concussion Assessment Tool (SCAT) is the most frequently utilized sports tool in the world.
The SCAT1 was developed in 2004 by combining the SAC, the Post-Concussion Symptom Scale (PCSS), sports-specific orientation questions, on-field concussion indications, and RTP recommendations~\cite{echemendia2017sport}.
After its appearance, SCAT has been redesigned and updated its versions by combining other assessments to improve the efficiency of the assessment.
To assist in the diagnosis of SRC, SCAT5 was created for qualified healthcare practitioners, usually including doctors and sports trainers. 
The SCAT5, which has undergone multiple significant revisions, is still the industry standard for sideline evaluation. 
Depending on the degree of the brain damage, administering SCAT5 might take as long as 20 minutes~\cite{echemendia2017sport}.

The above-described exams are extremely subjective in nature, and they take a long time. Other tests have been tried and shown to be helpful in identifying SRC on the sidelines at the time of injury in addition to the SCAT. The King-Devick Test (KD) detects acutely subtle visual scanning abnormalities and is based on the idea that mTBIs are a multi-system lesion. KD is a quick visual scanning test lasting 2 to 3 minutes in which participants read numbers from left to right for several lines on stimulation pages~\cite{galetta2016king}. 
KD is a potential indicator of mTBIs with good specificity and sensitivity levels. Combining KD with other popular rapid sideline assessments of focus, memory, and balance increases sensitivity and specificity~\cite{wright2017visual}.

Considering the other assessment methods, the VOMS has proven to be a clinically effective tool to diagnose concussion that recorded a high accuracy of 89\% with areas under the receiver operating characteristic curve (AUC) for adolescent and collegiate-athlete populations \cite{kontos2021discriminative}. Even VOMS manifested high internal consistency in civilian \cite{moran2018reliability} and military \cite{kontos2021discriminative} people, making it one of the most popular multi-domain assessments for the concussion detection system. Current sideline assessment tools SCAT3, SAC, KDT, BESS, etc. failed to inscribe vestibular and ocular system dysfunction comprehensively \cite{sussman2016clinical}. The VOMS allows incorporating the missing findings related to the vestibular and ocular system while assessing concussion. However, the VOMS assessment is also distinctly subjective as the patients verbally rate changes in different symptoms including headache, dizziness, nausea, and fogginess on a scale of 0 to 10 after each VOMS assessment in comparison to their pre-assessment state \cite{mucha2014brief} \cite{kontos2016reliability}. 

Because the VOMS relies on subjective reporting of provoked symptoms, it is important to understand how the eye movements may be related to these provoked symptoms.
In order to implement the concept and eliminate the subjectivity from the assessment, we developed a novel method for VOMS testing utilizing virtual reality.
In this paper, we analyze the smooth pursuit test of VOMS data collected through VR technology using different deep learning architectures.

\section{Methodology}
\subsection{Data Collection}

\subsubsection{Apparatus}

The Unity3D engine version 2019.1.6 and the SteamVR plugin version 1.7 were used to develop virtual VOMS stimuli through which the VOMS protocol was simulated in a VR environment. To display the VR stimulus, we used the HTC Vive Pro Eye Head Mounted Display (HMD) with a field of view (FOV) of 110$^o$, refresh rate of 90Hz, a combined resolution of 2880×1600 pixels, six degrees of freedom (DoF) for position and orientation tracking, and adjustable interpupillary (IPD) and focal distances. 
The VR HMD was powered by an Acer Predator gaming laptop with a 7th Generation Intel Core i7 Quad-Core processor with 16GB of memory and NVIDIA GeForce GTX 1070 graphics card running Windows 10. 



The integrated binocular eye tracker of the HTC Vive Pro Eye HMD were used to collect eye tracking data from participants. The eye tracker provides data at a sampling rate of 120 Hz, and has a trackable FOV of 110$^o$. Our application was equipped with procedures to collect eye tracking data at a 120Hz sampling rate along with head position and orientation data from the SteamVR SDK. HTC’s Sranipal SDK version 1.1.0.1 was used to read eye tracking data from the eye trackers. 

\subsubsection{Procedure}

After participants arrived, they were briefed about the experimental protocols and were instructed to be seated for the experiment.
The smooth pursuit experiment started by calibrating the eye trackers with a 5-point calibration procedure.
23 subjects, medically diagnosed with sport related concussion (SRC), were evaluated. 
As a part of the post-injury battery of assessments, the VOMS was administered via a virtual reality (VR) system and software package. 
Prior to administration of the VOMS in VR (VR-VOMS), baseline symptomology data was collected verbally that was specific to headache, nausea, dizziness, and fogginess. Subjects were asked to state the severity of each symptom based upon a scale of 0-10. 
After each of the 7 tests within the VOMS protocol, symptomology was rated by the subject and recorded. The symptom change score was then calculated as a measure of the baseline symptom scores compared to the overall change in symptom ratings over the 7 tests within the VOMS protocol~\cite{mucha2014brief}.

The smooth pursuit sub-test tests the ability of the participant to follow a slowly moving target stimuli. Our smooth pursuit test protocol used a 14 pt font-sized simulated target placed at a distance of 3ft from the participant's eyes. The participant is instructed to maintain focus on the target stimuli, as it moves smoothly, first 1.5 ft left and right in the horizontal direction and then 1.5ft up and down in the vertical direction.


\begin{figure}[t]

\begin{minipage}[b]{.45\linewidth}
  \centering
  \centerline{\includegraphics[width=4.5cm]{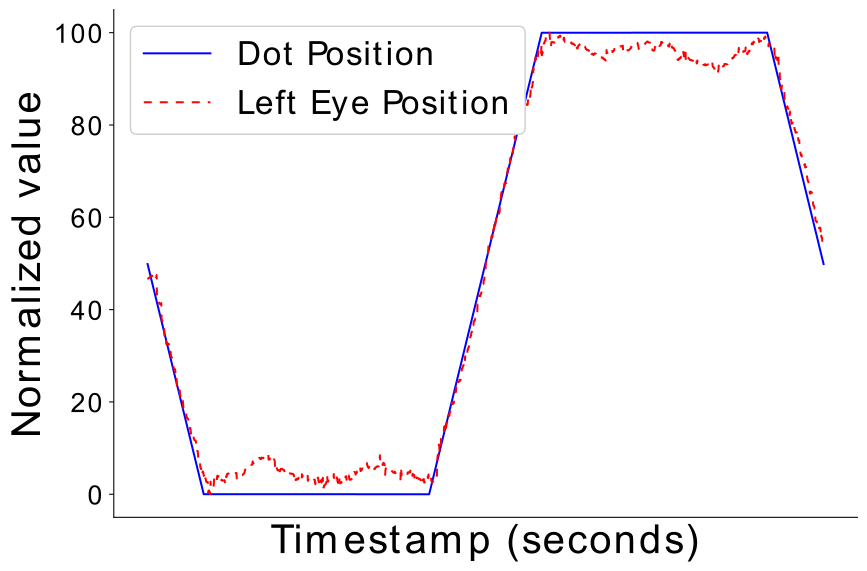}}
  \centerline{(a)}\medskip
\end{minipage}
\hfill
\begin{minipage}[b]{.45\linewidth}
  \centering
  \centerline{\includegraphics[width=4.5cm]{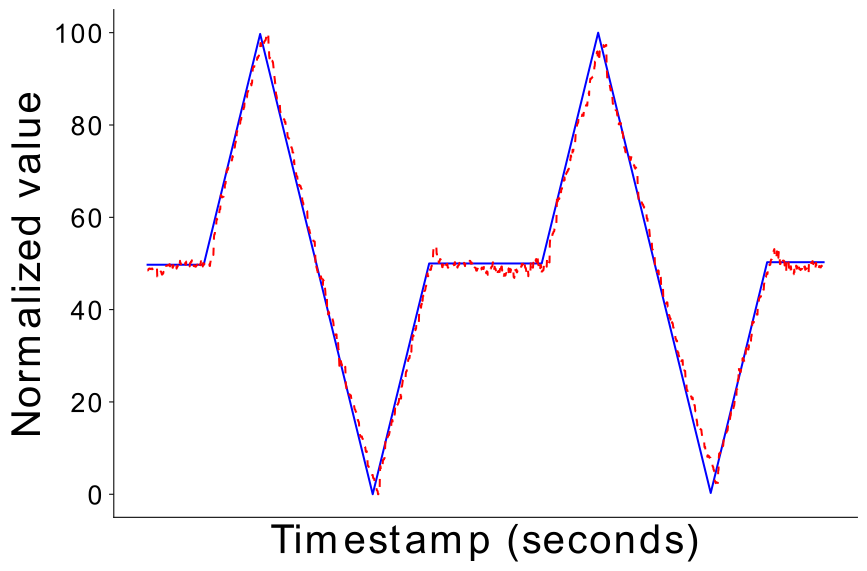}}
  \centerline{(b)}\medskip
\end{minipage}

\begin{minipage}[b]{.45\linewidth}
  \centering
  \centerline{\includegraphics[width=4.5cm]{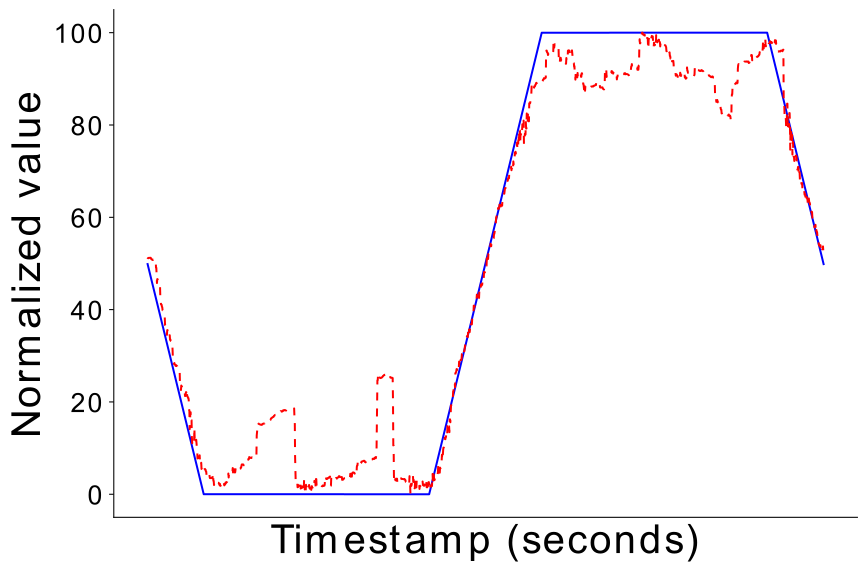}}
  \centerline{(c)}\medskip
\end{minipage}
\hfill
\begin{minipage}[b]{0.45\linewidth}
  \centering
  \centerline{\includegraphics[width=4.5cm]{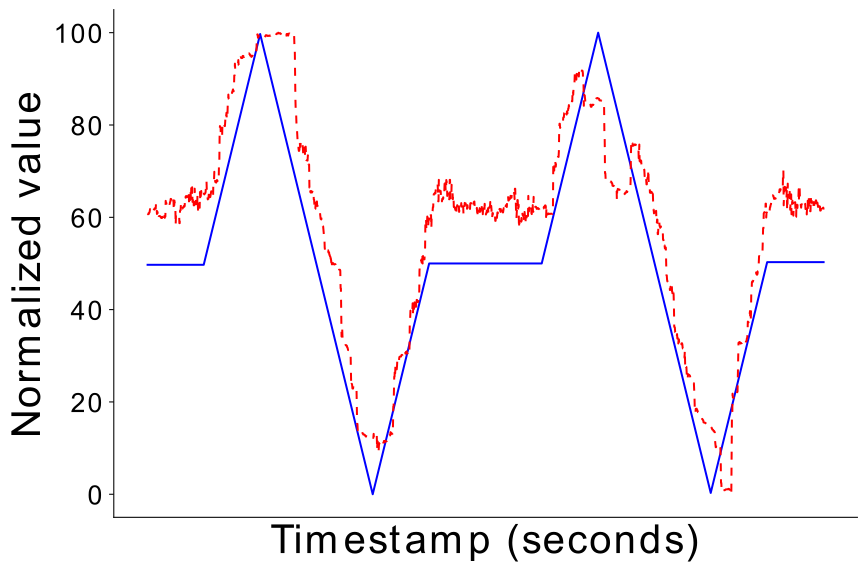}}
  \centerline{(d)}\medskip
\end{minipage}

\caption{Smooth pursuit normalized left eye signal and visual stimuli signal of x and y axis in control shown in (a) and (b) and SRC shown in (c) and (d) respectively vs timestamp in seconds.}
\label{fig:eyedotposition}

\end{figure}

\subsection{Network Description}

Recurrent neural networks (RNN) have been successful and effective for learning and predicting sequential data. RNNs with Long Short-Term Memory (LSTM) offer highly sophisticated techniques for sequential data encapsulating correlations between two points that are close in the sequence which is why they have been used in many state-of-the-art models to solve challenging problems \cite{greff2016lstm}. Since the dataset contains sequential data of gaze positions at different timestamps, the rational choice for this problem is using recurrent neural networks. Again, binary classification on the dataset leads to lower accuracy as the dataset does not contain enough data to approach this as a classification problem. So we encountered this as a regression problem. The details of the regression method and architecture are described in the next section.

\subsubsection{Regression Methodology}

Let, U be the set of all human signals in response to visual stimuli of VOMS assessment. The set consists of healthy or control and unhealthy or concussed patient signals namely $\varphi_{healthy}$ and $\varphi_{unhealthy}$ so that,
$$ \varphi_{healthy} + \varphi_{unhealthy} = U $$ 

Now, we use a deep learning model, $\psi$ to train so that it takes control signal as input and outputs the coordinates of the triggered visual stimuli to the patient, $y$. If x is a control signal, the model, $\psi$, is going to predict so that, 

$$P(x) \xrightarrow{\text{$\psi$}} \varphi$$

If any output of the given input patient signal from the model is $\hat{y}$, the loss function is mean squared error (MSE), $\mid \mid y - \hat{y} \mid \mid ^2$. For training, the model tries to minimize the loss so that the difference between $y$ and $\hat{y}$ is minimum for the control input signal. Based on the output of the loss function, a threshold value of probability is to be chosen which differentiates control signal from concussed patient or SRC signal.

Now, if an unknown signal $\bar{x}$ is input at the model, the model would try to predict the probability, $P(\bar{x}|\psi)$. If the $P(\bar{x}|\psi) < $ threshold value, the signal is to be considered as the control signal, $\varphi_{healthy}$. Otherwise, the input signal is to be considered as a SRC signal, $\varphi_{unhealthy}$.


\begin{figure}[t]
  \centering
  \centerline{\includegraphics[width=1.0\linewidth]{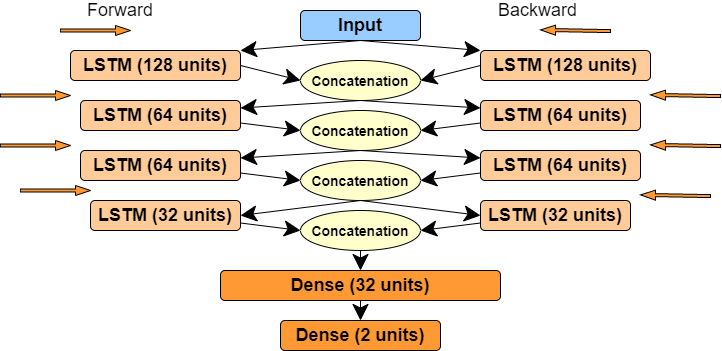}}
  \caption{Bidirectional LSTM network architecture}
  \label{fig:modelimage}
\end{figure}

\subsubsection{Architecture}
For the regression problem, we used several models to compare the results. All of the models are based on LSTM because of the sequential nature of the dataset. In the first two models, we used unidirectional vanilla LSTM and bidirectional LSTM \cite{schuster1997bidirectional}. Bidirectional LSTM consists of two LSTMs- forward and backward and it boosts the quantity of the data the network could access for more precise prediction by using the context of both previous and latter input data at a particular timestamp. The LSTM layers used in unidirectional LSTM are 128, 64, 64, 32 respectively, and finally 32 and 2 dense layers at the output. In the case of the bidirectional LSTM model, identical number of bidirectional LSTM and dense layers are used as shown in Fig 2. In the next model, we used residual LSTM \cite{wang2016recurrent} (of four 64 layers of LSTM) on top of bidirectional LSTM to increase the depth of the model and to overcome the degradation problem. In the final model, we used CNN LSTM \cite{wang2016dimensional} to capture better spatiotemporal correlations among the data points and used four 32 units of Conv1D in addition to the previously mentioned bidirectional LSTM architecture.

\subsection{Proposed Metric}
\label{ssec:proposedmetric}

To evaluate the performance of the models, we calculated the deviation of the prediction from the original patient signal for both x and y axis. In this paper, we propose a new error metric, MSE$^{\prime}$, for detecting concussion where the proportion of mean squared error of x and y axis with proportion 1 : 3. So, $MSE$ $^{\prime}$ = 0.25 $\times$ $MSE_x$ + 0.75 $\times$ $MSE_y$ where $MSE_x$ and $MSE_y$ are mean squared error of x any y signals respectively.

\section{Experiments}
\label{sec:experiment}

\subsection{Dataset and Preprocessing}
\label{ssec:datapreprocessing}

The VR-VOMS dataset contains eye gaze positions of 228 controls and 23 SRC from the smooth pursuit test of the VOMS assessment. Each patient excel file contains data of eye direction, eye position of x, y, and z axis for left, right and cyclopean eye with associated timestamps. It also contains the x, y, and z axis coordinates of the stimuli shown inside VR and details about their injury such as days before the test the patient got injured, scaling of their total VOMS change score. For training our model, we used gaze position (eye direction) data of x and y axis for left, right and cyclopean eyes as inputs and stimuli position inside VR as output. Fig 1 shows that the eye movement in x and y axis of healthy (control) and concussed (SRC) patient follows the dot position of x and y axis in VR which validates our choice of the gaze position columns as input. Before training the models, we normalized all columns of each patient data between 0 to 100 to have a common scale.

\subsection{Performance Metric}
\label{ssec:metric}
 In Table 1, we compared results of the proposed error metric ($MSE$ $^{\prime}$) with mean squared error of x and y axis with proportion 1 : 1 ($MSE$ = 0.5 $\times$ $MSE_x$ + 0.5 $\times$ $MSE_y$). The optimal threshold for each model is chosen in such a manner that the threshold maximizes $(TPR - FPR)$, where TPR and FPR is true positive rate and false positive rate respectively. Each error metric is then compared with three performance metrics namely accuracy, sensitivity and specificity. 
 
\begin{equation}
    \label{eq-acc}
    \text{Accuracy} = \frac{1}{N} \sum (TP + TN)
\end{equation}

\begin{equation}
    \label{eq-sens}
    \text{Sensitivity} = \frac{TP}{TP + FN}
\end{equation}

\begin{equation}
    \label{eq-spec}
    \text{Specificity} = \frac{TN}{TN + FP}
\end{equation}

where $TP$, $FN$, $TN$ and $FP$ are true positive, false negative, true negative and false positive respectively.

\begin{figure*}[t]
 
\begin{minipage}[b]{0.225\textwidth}
  \centering
  \centerline{\includegraphics[width=1.0\linewidth]{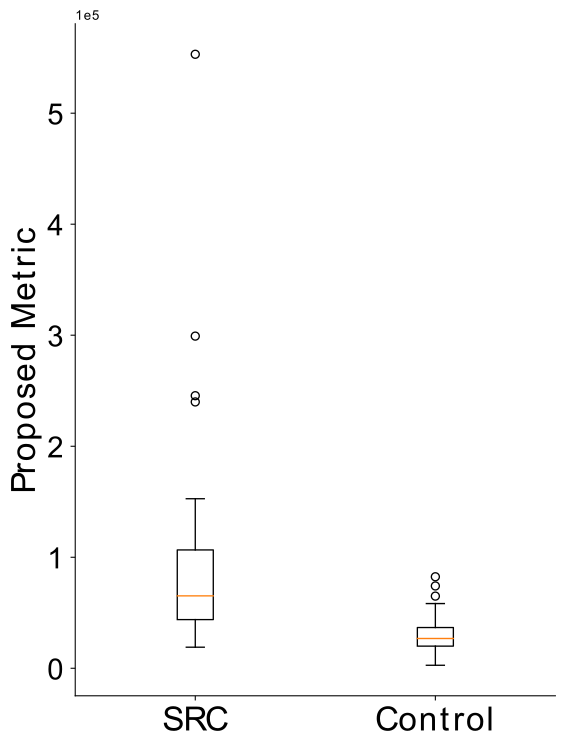}}
  \centerline{(a)}
\end{minipage}
\hfill
\begin{minipage}[b]{0.75\textwidth}  
  \centering
  \centerline{\includegraphics[width=1.0\linewidth]{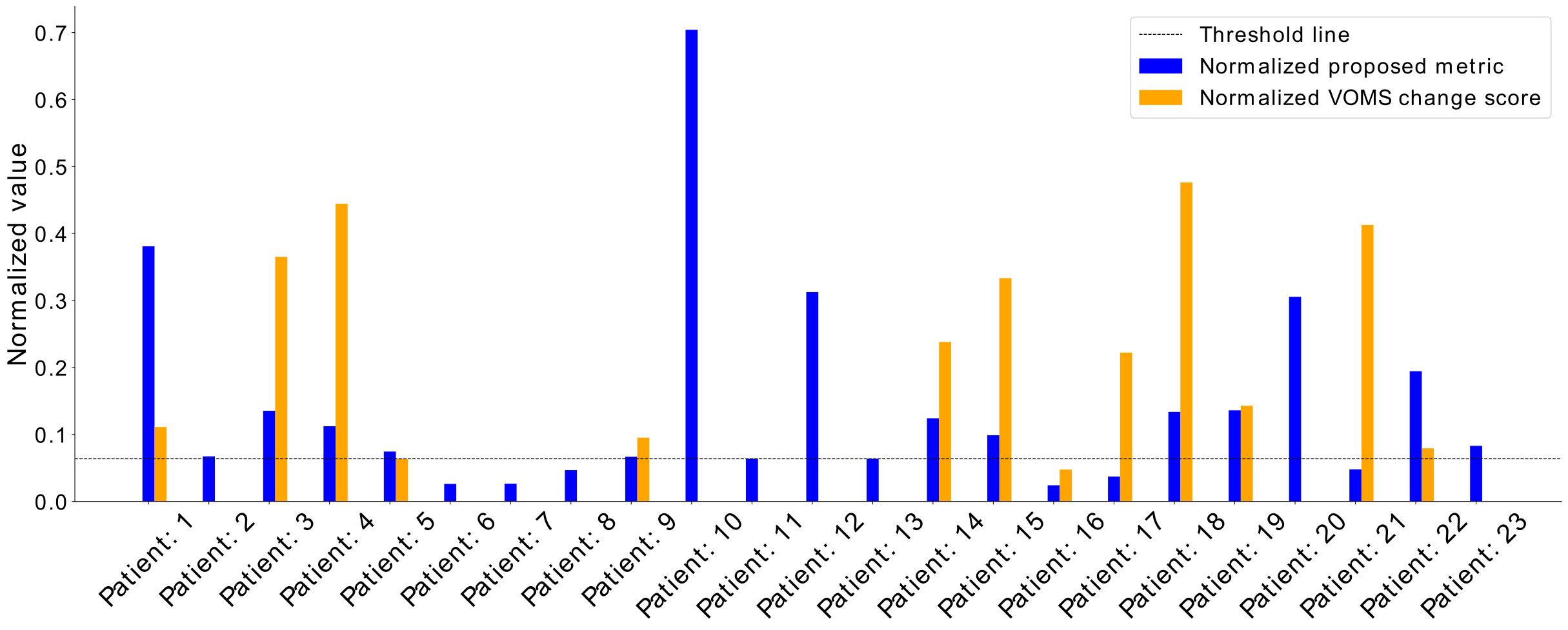}}
  \centerline{(b)}
\end{minipage}

\caption{Comparison of our proposed metric ($MSE^{\prime}$) between SRC and control using box plot in (a) and comparison of proposed metric in smooth pursuit test with respect to overall VOMS change score for SRC in (b).}
\label{fig:VOMSComparison}
\end{figure*}

\begin{table}[ht]
\centering
\small
\caption{Test Results on VR-VOMS Dataset}
\medskip
\setlength{\tabcolsep}{1pt}
\begin{tabular*}{1.0\columnwidth}{@{\extracolsep{\fill}}c c c c c}

    \toprule
    \textbf{Architecture}
    & \multicolumn{1}{p{1.4cm}}{\centering \textbf{Metric}}
    & \multicolumn{1}{p{1.4cm}}{\centering \textbf{Accuracy}  \\ \textbf{(\%)}}
    & \multicolumn{1}{p{1.6cm}}{\centering \textbf{Sensitivity}  \\ \textbf{(\%)}}
    & \multicolumn{1}{p{1.6cm}}{\centering \textbf{Specificity}  \\ \textbf{(\%)}} \\ \midrule
    
    Unidirectional  & MSE$^{\prime}$    & 90.84             & 73.91             & 92.54 \\
    LSTM            & MSE               & 84.86             & 73.91             & 85.96 \\          \midrule
    
    Bidirectional   & MSE$^{\prime}$    & \textbf{92.43}    & \textbf{73.91}    & \textbf{94.29} \\ 
    LSTM            & MSE               & 84.86             & 73.91             & 85.96  \\         \midrule
      
    Residual LSTM   & MSE$^{\prime}$    & 88.05             & 65.22             & 90.35 \\
                    & MSE               & 78.49             & 73.91             & 78.95 \\          \midrule
      
    CNN LSTM        & MSE$^{\prime}$    & 52.99             & 30.43             & 55.26  \\
                    & MSE               & 52.59             & 30.43             & 54.82 \\          \bottomrule
\end{tabular*}
\end{table}

\subsection{Hyper-parameter Tuning}
\label{ssec:hyperparameter}
For training the models, we used the Adam optimizer for all of the models with default learning rate of 0.001 in Tensorflow library. If the validation loss did not decrease for 3 consecutive epochs, the learning rate was reduced by factor of 0.1. Mean squared error was chosen as the loss function for training all models. Batch size and epochs for training the models was 256 and 40 respectively.

\section{Results \& Discussion}
\label{sec:results}

It is evident from Table 1 that using our proposed metric ($MSE^{\prime}$), the bidirectional LSTM model can detect concussion with over 92 percent accuracy. Bidirectional LSTM model increased the amount of information available to the network at a particular time to predict more accurately compared to unidirectional LSTM, residual LSTM and CNN-LSTM networks by taking input in forward and backward direction as shown in Fig 2. The results also justify our choice of metric as $MSE^{\prime}$ over threshold outperforms $MSE$ by more than 7.5\% accuracy for bidirectional LSTM. This indicates that the deviation of gaze from y-axis has more prominence than the deviation from x-axis for concussion. In Fig 3(a), we compared the proposed error metric for SRC and control using a box plot which shows the median of the error metric for SRC is higher than that of the control. Again, as shown in Fig 3(a), there is an overlap of region between SRC and control for the proposed metric and the model predicted false positive and false negative for patient data in this region. 

From Fig 3(b), when the proposed metric value for any patient is above the threshold line, our model classifies them as concussed patients. Additionally, we compared our proposed metric with VOMS change score which is the total symptoms (verbally rated) provoked during the test. Our model is efficient enough to predict when the patient did not show any symptom during the manual VOMS assessment as shown in Fig 3(b). This also demonstrates the subjective nature of the manual VOMS assessment and our approach is able to develop objectivity in concussion detection. Also, Few SRC patients were able to perform the smooth pursuit test without any trouble. Our model could not detect any concussion symptoms in those cases which lead to low sensitivity. Again, low specificity in test results indicates that either some controls had concussion symptoms or they did not follow the experiment protocol properly making their gaze position signal deviated from the stimuli. In Table 1, it is noticeable that some models have same sensitivity which demonstrates that even though those models could predict same true positives, they optimized in distinctive manner under specified condition which lead to different sensitivity.

\section{Conclusion and Future Work}
\label{sec:conclution}

In this paper, we propose a novel approach to detect concussion symptoms by incorporating bidirectional LSTM and a new metric. Moreover, by comparison with traditional VOMS assessment, we demonstrated that our method can classify potential concussion with higher accuracy. In future, we wish to extend our current research by analyzing other aspects of the VOMS assessment- saccades, convergence, VOR, VMS and construct a more robust method for detection of concussion accurately. Finally, our goal is to identify other features of both vestibular and ocular behavior of the assessment, so that it can be used by athletic trainers and clinicians for trouble-free, yet valid and precise diagnosis of concussion.

\subsubsection{Acknowledgements} 
Portions of this material is based upon work supported by the Office of the Under Secretary of Defense for Research and Engineering under award number FA9550-21-1-0207.

\bibliographystyle{ieeetr}
\bibliography{bibliography.bib}
\end{document}